\documentclass[sigplan,nonacm]{acmart}

\usepackage{amsmath}
\usepackage{booktabs}
\usepackage{graphicx}
\usepackage{dblfloatfix}
\usepackage{multirow}
\usepackage{xcolor}
\usepackage{xspace}

\newcommand{\sys}{KVSeal\allowbreak\xspace}
\newcommand{\placeholderfigure}[2]{%
  \fbox{\parbox[c][#1][c]{0.94\columnwidth}{\centering\small #2}}%
}
\newcommand{\runinheading}[1]{%
  \par\noindent\hspace*{\parindent}%
  {\normalfont\bfseries #1.}\enspace%
}

\newcommand{\preprintpagestyle}{%
  \fancyhf{}%
  \renewcommand{\headrulewidth}{0pt}%
  \renewcommand{\footrulewidth}{0pt}%
  \fancyfoot[C]{\footnotesize\thepage}%
}
\AtBeginDocument{%
  \fancypagestyle{standardpagestyle}{\preprintpagestyle}%
  \fancypagestyle{firstpagestyle}{\preprintpagestyle}%
  \pagestyle{standardpagestyle}%
}

\newcommand{\preprintinstitute}{Institute of Parallel and Distributed Systems}
\newcommand{\preprintuniversity}[1][J]{Shanghai #1iao Tong University}

\newcommand{\slantedJ}{%
  \hbox{\pdfsave\pdfsetmatrix{1 0 0.20 1}%
    \rlap{\textup{J}}\pdfrestore\phantom{\textup{J}}}%
}

\makeatletter
\renewcommand{\@mkauthors}{%
  \begingroup
  \nxandlist{, }{, }{, }\authors
  \global\setbox\mktitle@bx=\vbox{%
    \hsize=\textwidth
    \unvbox\mktitle@bx
    \medskip
    \centering
    {\Large\authors\par}
    \smallskip
    \textit{\preprintinstitute}\par
    \textit{\preprintuniversity[\slantedJ]}\par
    \bigskip
  }%
  \endgroup
}
\makeatother

\begin{document}

\title{\sys: Serving Long-Context LLMs with Losslessly Compressed KV Caches}

\author{Hangyu Liang}
\author{Zeyu Mi}
\affiliation{%
  \department{\preprintinstitute}
  \institution{\preprintuniversity}
}

\begin{abstract}
Long context is what lets an LLM serve real workloads: a single request can
carry an entire repository, a document collection, or a long agent trajectory.
It also makes the KV Cache, rather than the model weights, the dominant
serving cost, because the cache grows linearly with the context and every
decode step rereads all of it. Eviction, sparse attention, and low-bit
quantization shrink this cost by changing which K/V entries attention sees or
what values they carry, so accuracy that holds on short prompts can collapse
on the long-context tasks these systems target. Lossless compression can
reduce KV storage while preserving every original K/V bit. Prior uses of
lossless compression include shrinking static weights and reducing KV
transfers between workers. Applying it to resident KV, however, requires
the serving system to compress pages online as they fill, manage them
independently as requests progress, and keep them accessible to decode
attention.

We present \sys, which makes lossless compression the resident form of the KV
Cache. We exploit the strong concentration of BF16 KV exponents to design a
lossless page format with fixed-length codes and bounded escape storage.
Each compressed page carries its own encoded values and escape records,
allowing it to be stored, reconstructed bit for bit, and consumed by
attention independently of other pages. A compression-aware page lifecycle manages
raw and compressed pages as the cache grows. During decode, attention reads
both formats and reconstructs compressed values as it consumes them.
\sys achieves about \(1.3\times\) resident KV compression, up to
\(1.27\times\) decode attention speedup over FlashAttention-2, and up to
\(1.19\times\) end-to-end decode speedup over stock vLLM.

\end{abstract}

\maketitle

\section{Introduction}
\label{sec:introduction}

Long-context inference is becoming a default interface for frontier LLM
services. Representative offerings from OpenAI, Anthropic, Google, Z.AI, and
Moonshot now advertise context windows of approximately one million
tokens~\cite{openai-gpt6-astra,anthropic-fable51,google-gemini38,zai-glm52,kimi-k3}.
Yet this capability turns context length into a serving-system cost: every
generated token appends K/V vectors, while each decode step repeatedly reads
the growing history. The resulting KV Cache grows linearly with sequence
length. On Llama 3.1 8B, for example, a single 256K-token sequence requires
about 32~GiB of BF16 K/V storage, more than twice its 8B BF16
parameters~\cite{llama31,pagedattention}. Decode attention scales with the
same quantity, because a one-token query must load and process the entire
cached history; FlashAttention-2 (FA2) identifies this KV-cache loading,
rather than the intermediate attention matrices, as the bottleneck in this
regime~\cite{flashattention2}. Long-context serving must therefore reduce both
the GPU memory held by resident KV and the K/V traffic each decode step
incurs.

Existing KV-cache optimizations trade memory and bandwidth savings against
model fidelity. Quantization methods, such as TurboQuant, VecInfer, and KV
Cache Transform Coding, save memory by changing the numerical representation
of K/V values~\cite{turboquant,vecinfer,kv-transform-coding}. Sparse attention
methods, including Squeezed Attention, RetrievalAttention, and SolidAttention,
reduce memory traffic and attention cost by restricting the history examined
by each query~\cite{squeezedattention,retrievalattention,solidattention}. Both
families save substantial resources, but neither can tell in advance whether a
coarsened value or a skipped entry is the one a later query will need. Such
approximations may be adequate at short context, where the evidence for an
answer is rarely far away, yet on long-context tasks the altered or omitted
entries can cost substantial accuracy~\cite{kv-transform-coding}.

Lossless compression offers a compelling way to reduce KV memory while
preserving bit-exact recovery whenever the stored values are needed. Prior work
has explored this direction for different targets: ZipServ~\cite{zipserv}
compresses static model weights, SplitZip~\cite{splitzip} compresses KV for
disaggregated transfer and reconstructs it at the receiver, while Kamath
et al.~\cite{kamath2026lossless} study lossless GPU-memory compression
for tensors including LLM KV caches. However, these approaches organize
compression around fixed tensors, tensor collections, or transfer objects,
rather than a resident KV Cache whose state changes as requests grow. Adapting
them to online KV serving raises three challenges.

\textbf{Challenge 1: Dynamic KV-cache lifecycle.} A resident KV cache grows as
requests append tokens, while existing pages may be reclaimed or change state
during serving. ZipServ pools exceptional values in globally indexed
buffers~\cite{zipserv}, while Kamath et al.~\cite{kamath2026lossless} exploit bit
patterns shared across tensors. In both designs, recovery information spans
more than one page. If pages with different lifetimes share that information,
the runtime must keep it available until the last page that depends on it is
reclaimed. Online serving therefore needs to compress, move, recover, and
reclaim each page independently while keeping other live pages accessible
and recoverable. In other words, each cache page must carry all of its
page-specific recovery data, so those data can move with the page and be
reclaimed with it, avoiding the need for a global recovery buffer as the
cache changes.

\textbf{Challenge 2: Coexistence of compressed and uncompressed pages.} A newly
created KV page may still be receiving tokens, so it is not yet suitable for
compression. Some completed pages are also unsuitable for compression
and remain in their original format.
Consequently, a request can contain both uncompressed and compressed pages,
with the mix changing as generation proceeds. The serving system must manage
their different physical representations without changing the logical KV
sequence. Attention must then consume both forms for the same query, rather
than requiring the entire cache to use one format.

\textbf{Challenge 3: Efficient consumption of compressed KV.} During decode,
attention repeatedly reads the resident KV Cache, making data movement a major
cost. A separate decompression step would write recovered BF16 pages to GPU
memory, only for attention to read them again. This extra traffic can outweigh
the bytes saved by compression. The challenge is to let attention consume
compressed pages without materializing a full BF16 cache.

To address these challenges, we present \textbf{\sys}, a serving system that
integrates lossless compression into the lifecycle and attention path of a
resident KV Cache. Its \textbf{page-local lossless format} pairs compact
exponent codes with a bounded region for escape records. The fixed layout
gives compressed pages of the same configuration a uniform slot size.
Given the calibrated exponent window, each page can be recovered from its
own data. Keys and values use separate layouts that match how decode
attention reads them.

To manage a cache that continues to grow during serving, \sys introduces a
\textbf{compression-aware page lifecycle}: it attempts compression when a page
becomes complete, while pages still being filled or unable to fit the
compressed representation remain raw. Both forms share the same logical page
interface, so their physical differences need not affect how the serving
system tracks a request's KV sequence. Finally, \sys uses
\textbf{mixed-format fused attention} to consume raw and compressed pages
together. Separate paths handle the two representations and combine their
attention results, while compressed K/V is reconstructed as it is consumed.
This avoids materializing a full BF16 cache, reduces KV data movement, and
turns the storage savings of lossless compression into faster decode attention.

We implement \sys by extending FlashAttention-2 kernels~\cite{flashattention2},
integrate it into vLLM~\cite{pagedattention}, and evaluate it on NVIDIA RTX 4090
and L40S GPUs.
The page-local format provides up to \(1.3\times\) physical KV compression
with bit-exact recovery of the original BF16 K/V values. Its encode and decode
throughput is also higher than that of the compared lossless codecs, keeping
the representation cheap to produce and consume. Fused attention achieves up
to a \(1.27\times\) speedup over FA2, and end-to-end decode achieves up to a
\(1.19\times\) speedup over stock vLLM.

In summary, our contributions are:

\begin{itemize}
  \item We design a lossless format for resident KV pages with fixed-size
  slots and bounded page-local escape storage. It supports independent,
  bit-exact recovery and arranges K and V for their respective attention
  access patterns.
  \item We develop a compression-aware residency model that manages raw and
  compressed pages throughout KV-cache growth, sealing, reuse, and reclamation.
  \item We develop mixed-format decode attention that reconstructs compressed
  K/V as it is read and merges partial attention results from raw and
  compressed pages under a shared softmax normalization.
\end{itemize}

\section{Background}
\label{sec:background}

\subsection{Attention and the KV Cache}
\label{sec:background-attention}

Transformer scaled dot-product self-attention~\cite{vaswani2017attention}
normalizes the similarity
between the current query and historical keys before aggregating the
corresponding values. For a decode token, we write
\[
o=\operatorname{softmax}\!\left(\frac{qK^{\mathsf T}}{\sqrt{d}}\right)V,
\]
where $K$ and $V$ denote the historical keys and values cached for the
current request. During prefill stage, the model processes many input tokens and
produces their K/V activations. During decode stage, each sequence appends one new
token per step but repeatedly reads the K/V of all previous tokens; caching
these activations avoids recomputing their projections. The
resident KV Cache therefore grows approximately linearly with context length,
and longer contexts increase both cache occupancy and the amount of historical
K/V read by attention. FlashAttention uses tiling and online softmax to avoid
materializing the full attention matrix and reduce intermediate traffic
~\cite{flashattention,flashattention2}; historical K/V nevertheless remains
resident data that attention must consume.

\subsection{BF16 Numerical Representation}
\label{sec:background-bf16}

BF16 is a 16-bit floating-point format widely used for large-model training
and inference. It consists of one sign bit, eight exponent bits, and seven
mantissa bits. For a normalized value, its numerical value can be written as
\[
x=(-1)^{\mathrm{sign}}\times
  2^{\mathrm{exponent}-127}\times(1.\mathrm{mantissa}).
\]
Compared with FP32, BF16 retains a similar dynamic range while halving the
storage cost of each element, which makes it common for model weights,
intermediate activations, and inference-time KV Caches. Prior work observes
that BF16 tensor exponents are not necessarily distributed uniformly over
the full 8-bit range; Huffman- or ANS-style entropy coding, as well as
GPU-oriented fixed-length codes, can exploit this statistical redundancy
to compress BF16 tensors~\cite{zipserv,heilper2025lossless,zipnn,dlf}.

\subsection{Lossless Compression Formats and GPU Constraints}
\label{sec:background-gpu}

Exploiting tensor redundancy during inference must respect the GPU execution
model. GPU kernels typically execute in SIMT: many threads apply the same
instructions to different data and obtain throughput through coalesced
global-memory accesses and bounded register and shared-memory working sets.
A GPU-friendly data path should therefore allow a thread to compute an
access address from the logical position while keeping control flow and work
per thread reasonably uniform. Data-dependent serial parsing, divergent
execution lengths, and frequent accesses to irregular metadata all reduce
parallel efficiency and add memory overhead. In a Huffman-like variable-length
bitstream, for example, symbol boundaries depend on preceding codeword
lengths, so a thread cannot determine the address of a target element from
its logical position alone. Parallel lookup then requires additional chunk
offsets, prefix indices, or coordination, while different threads may take
different numbers of decoding steps; these effects can cause warp divergence,
non-coalesced accesses, and extra metadata reads~\cite{dietgpu,recoil}.

The GPU data path is also poorly matched to a serial ``decompress everything,
then compute'' pipeline. If compressed data is first restored into a complete
BF16 buffer and written to global memory, the subsequent computation must read
that buffer again; the extra write and reread can offset the HBM traffic
saved by compression. Prior GPU compilers and kernel-fusion systems reduce
intermediate data movement through operator fusion and tile-level
reuse~\cite{welder,bolt,mirage}. Accordingly, compressed values should be
reconstructed as they are consumed and kept within the on-chip workspace
reusable by the current compute tile.

\subsection{Key Observations for Lossless KV Serving}
\label{sec:background-observations}

\begin{figure}[tbp]
  \centering
  \IfFileExists{figures/global_exponent_top16_coverage.pdf}{%
    \includegraphics[width=\columnwidth]{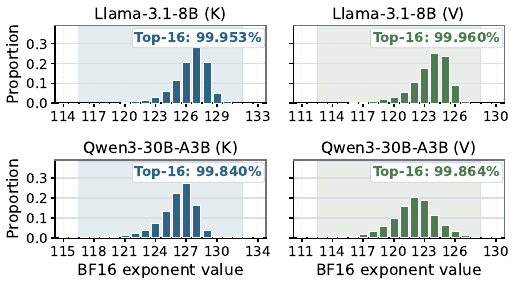}%
  }{%
    \placeholderfigure{1.45in}{Global BF16 KV exponent distribution and
    Top-16 coverage (figure pending).}%
  }
  \caption{Global BF16 exponent distributions of K/V activations.}
  \Description{Four panels show BF16 exponent distributions for Keys and
  Values in Llama-3.1-8B-Instruct and Qwen3-30B-A3B-Instruct-2507, with a
  global contiguous Top-16 interval highlighted after aggregating all layers
  and KV heads.}
  \label{fig:top16}
\end{figure}

\runinheading{Observation 1: Concentrated BF16 Exponents}
\label{sec:background-observation-exponent}
The core opportunity for KV compression comes from statistical redundancy in
BF16 exponents. We collect KV Cache data from Llama-3.1-8B-Instruct and
Qwen3-30B-A3B-Instruct-2507 on LongBench and PG-19, and count BF16 exponents
separately for Keys and Values. Figure~\ref{fig:top16} shows the resulting
global distributions, with values from all layers and KV heads aggregated
within each K/V type. It reveals a pronounced concentration in the
exponent distributions: in most model/tensor distributions, the high-probability
region forms a contiguous window of 16 exponent values and covers
99.84--99.96\% of all values. The tail still varies across models and K/V
types, but most KV values have exponents within a relatively small contiguous
range.

This result indicates that most exponents in real KV activations do not need
the full 8-bit range, leaving room for a more compact representation that
reduces KV storage and data movement. The reported coverage is a global
aggregate; separating the statistics by layer, KV head, and K/V type allows
each group to use its own exponent window and therefore achieves no lower,
and often higher, coverage.

\runinheading{Observation 2: Bandwidth-Bound Decode Attention}
\label{sec:background-observation-bandwidth}
Decode attention is often constrained by the cost of moving the KV Cache rather
than by arithmetic. For each request, one decode step contributes a single
query, yet dense attention must read the request's accumulated K/V history to
compute its output. The attention work per byte transferred is therefore
small, giving decode attention low arithmetic intensity and making memory
bandwidth a primary performance bottleneck~\cite{roofline}. This bandwidth-bound
regime also changes how decompression overhead should be viewed: lightweight
decompression need not be added serially to memory-fetch time. In a pipelined
execution, decompression of one K/V tile can overlap with loading another,
hiding part of its latency behind data movement. The relevant overhead is
therefore the portion of decompression that remains exposed after overlap; if
it is smaller than the time saved by reading fewer K/V bytes, lossless
compression can improve decode performance.

\runinheading{Observation 3: Context-Amplified Attention Cost}
\label{sec:background-observation-long-context}
For a fixed number of newly processed tokens, position-wise operators such as
FFNs perform roughly the same amount of work regardless of the preceding
context length. Attention differs because each new query must be evaluated
against the K/V context, so its work depends on both the number of new queries
and the context they attend to. This distinction applies to both prefill and
decode: prefill processes multiple new queries, whereas decode processes one
query per sequence per step. As context grows, attention work can therefore
increase without a proportional increase in FFN work, potentially shifting a
larger share of runtime toward attention.

\begin{figure}[tbp]
  \centering
  \IfFileExists{figures/attention_share_profile.pdf}{%
    \includegraphics[width=\columnwidth]{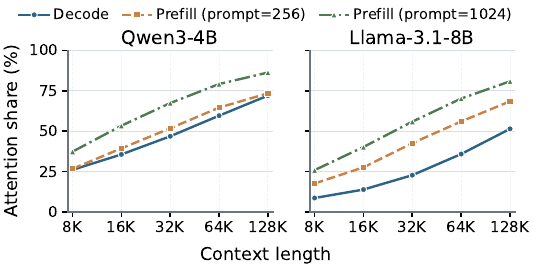}%
  }{%
    \placeholderfigure{1.35in}{Attention-core share versus cached context
    length for decode and prefill (figure pending).}%
  }
  \caption{Attention-core share versus cached context length.}
  \Description{Two side-by-side line plots show attention-core time share
  versus cached context length for decode and prefill with 256 or 1024 new
  queries in Llama-3.1-8B and Qwen3-4B.}
  \label{fig:attention-share}
\end{figure}

We quantify this effect with paired FA2 profiles on an
NVIDIA L40S using BF16, \(\mathrm{batch}=1\), and TP=1 for Llama-3.1-8B and
Qwen3-4B. Figure~\ref{fig:attention-share}
reports the attention-core time as a fraction of the corresponding phase time
over increasing cached histories. Each panel includes decode, which adds one
new query, and prefill with 256 or 1024 new queries. The attention share grows
with context length in both phases, and larger prefill batches make the
attention core a larger fraction of the phase time.
For a given fractional reduction in attention time, the larger decode
attention share at longer contexts leaves more room to reduce per-token
latency. The serving experiments measure how much of that opportunity
\sys realizes.

\section{System Design}
\label{sec:design}

\begin{figure}[tbp]
  \centering
  \IfFileExists{figures/kv_lossless_system_overview.pdf}{%
    \includegraphics[width=\columnwidth]{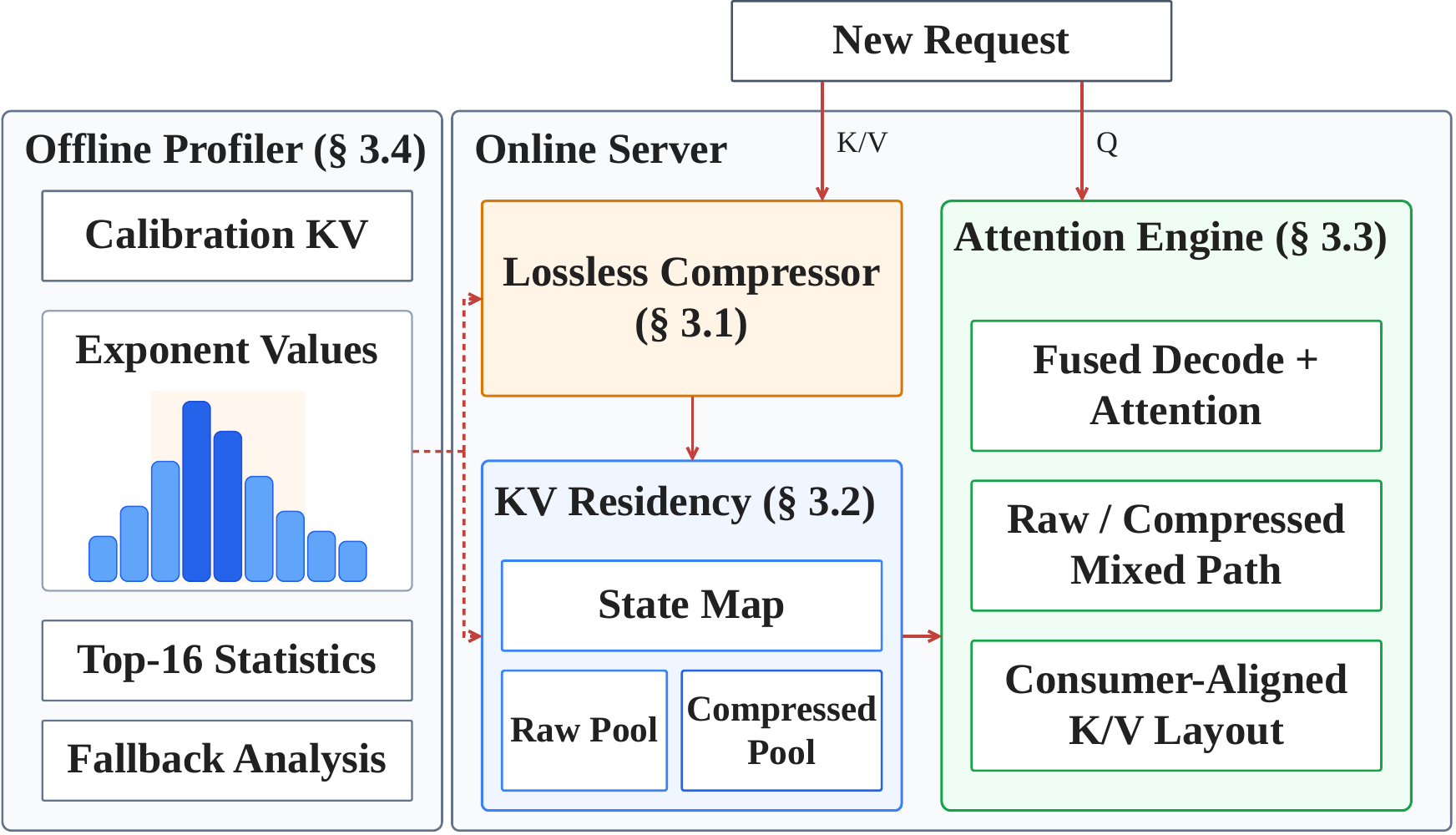}%
  }{%
    \placeholderfigure{2.0in}{System overview: offline calibration feeds
    online compression, KV residency, and attention execution.}%
  }
  \caption{Overview of \sys.}
  \Description{The figure separates an offline profiler from an online server.
  The profiler calibrates KV exponents and escape-count statistics. A new request
  enters the online server, whose lossless compressor feeds a residency layer
  with raw and compressed pools; the attention engine reads that state through
  fused attention paths, a raw/compressed mixed path, and consumer-aligned K/V
  layouts.}
  \label{fig:overview}
\end{figure}

Figure~\ref{fig:overview} summarizes the two parts of \sys: offline profiling
and online serving. The Offline Profiler
(\S~\ref{sec:design-calibration}) uses calibration data to make runtime
configurations. It selects model-specific exponent windows,
determines the capacities of the Raw Pool and Compressed Pool, and sets the
size of each slot. Here, a \textbf{slot} denotes a configured physical
allocation for page data, whereas a page is the logical unit identified by its
token range. \sys organizes the online path around page sealing. The Online
Server handles requests as they arrive: as a request generates tokens, its K/V
values are appended to the cache as pages grow. Once a page becomes full, it
seals and the Lossless Compressor
(\S~\ref{sec:design-format}) encodes it in the page-local lossless format. KV
Residency (\S~\ref{sec:design-residency}) manages page states and places raw
or compressed pages in their physical pools, allowing different
representations to coexist. The Attention Engine
(\S~\ref{sec:design-attention}) consumes these pages through fused compressed
attention or a mixed path, reconstructing compressed values as part of
attention rather than materializing a separate BF16 cache.

Overall, \sys reduces resident KV storage by keeping completed pages compressed
and lowers the KV traffic of decode attention by consuming those pages without
materializing a full BF16 cache. The reduced traffic can accelerate
bandwidth-bound decode. As Observation~3 shows, attention accounts for a larger
share of inference time with longer contexts, making these benefits particularly
relevant to long-context serving.

\subsection{Lossless Page Format and Recovery}
\label{sec:design-format}

\runinheading{4-bit Exponent Coding and Escape Records}
Observation~1 shows that KV
exponents typically concentrate in a short contiguous interval. We therefore
use fixed-length 4-bit exponent codes with sparse escape records. For
each layer, KV head, and K/V type, \sys selects a contiguous 16-exponent
window \( [b,b+15] \) and encodes it as \texttt{0000} through \texttt{1111}.
Here, \(b\) is the offline-calibrated base exponent loaded with the runtime
configuration; we discuss window selection and calibration in
Section~\ref{sec:design-calibration}. For an in-window value,
\(b+\text{code}\) exactly recovers the original exponent. An exponent outside
the window cannot be represented by these codes; we call such a value an
escape value. \sys stores one escape record for each escape value. Each record
occupies four bytes: 2 bytes record the value's page-local position and
another 2 bytes store its original BF16 bits. To keep every position on the
fixed-length path, the encoder writes code 0 for an escape value. The
resulting exponent \(b+0\) is only a placeholder and does not represent the
true exponent. The GPU can first reconstruct all positions in parallel and
then overwrite the placeholders using the positions and original bits in the
escape records. This design combines a fixed-length representation for common
values with exact storage for the few exceptional values, without trading
accuracy for compression.

\runinheading{Head-Page Organization}
KV Cache differs fundamentally from static model weights in its lifecycle.
Static weights can be scanned and organized as one continuous compressed
object before inference; for example, ZipServ designs a regularized compressed
layout for fixed weight tensors~\cite{zipserv}. KV Cache, however, is produced token by
token during serving, and existing pages may be shared, reclaimed, or reused.
Chaining pages into a cross-page variable-length global stream would make
insertion or release change the offsets and locations of subsequent data and
would require state spanning the entire cache. \sys therefore does not rely on
a global stream; it stores the data required for recovery in each page, while
static configuration such as the base exponent is supplied globally by the
runtime.

Inference frameworks already manage consecutive KV tokens in fixed-size blocks
or pages to support allocation, sharing, and reclamation~\cite{pagedattention}.
Let \(T\) denote the number of tokens per page and \(D\) the head dimension.
We call the K or V values of one KV head over a page's token range a
\emph{head-page}; it is the basic unit of compression and recovery. A
head-page contains \(N=TD\) BF16 values and has a raw size
\[
S_{\mathrm{raw}}(T,D)=2TD\ \text{bytes}.
\]

\begin{figure}[tbp]
  \centering
  \IfFileExists{figures/head_page_lossless_format.pdf}{%
    \includegraphics[width=0.9\columnwidth]{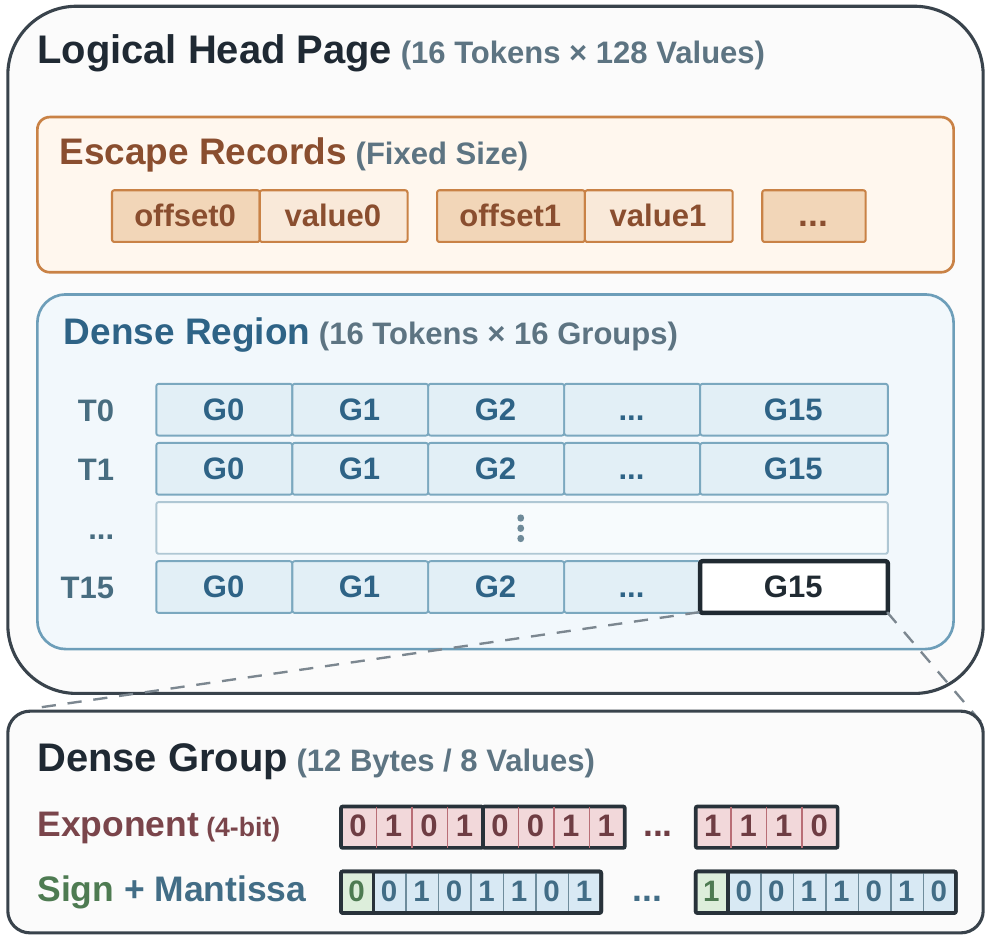}%
  }{%
    \placeholderfigure{1.8in}{Head-page format: dense groups plus page-local
    escape records.}%
  }
  \caption{Example head-page layout with 16 tokens and head dimension 128.}
  \Description{The figure shows a head-page with a fixed dense region and a
  page-local escape-record region. The expanded group uses one cell for each
  exponent bit and sign/mantissa bit, with thicker boundaries separating
  values and groups.}
  \label{fig:head-page-format}
\end{figure}

Figure~\ref{fig:head-page-format} shows an example head-page layout with
\(T=16\) and \(D=128\). Each head-page contains a fixed Dense Region and a
bounded Escape Records region. The Dense Region uses a fixed layout of
eight-value groups: each group packs eight 4-bit exponent codes into one
32-bit word and stores the corresponding sign/mantissa bytes in two 32-bit
words. Each group occupies 12 bytes, compared with 16 bytes in raw BF16 form,
giving an ideal compression ratio of \(1.33\times\); groups can be processed
independently by parallel threads. The K and V Dense Regions use layouts
tailored to QK and PV accesses, respectively. These layouts alter physical
placement without changing the logical token positions of recovered values.
Out-of-window values are retained in page-local Escape Records containing
their positions and original BF16 bits. Their Dense Region codes serve only as
placeholders and are overwritten during recovery, so each element remains
addressable at a fixed position. To keep physical slot sizes uniform among
compressed pages of the same type, the escape capacity is fixed; every
compressed page consequently reserves the same amount of record space, even
when it contains fewer escape records.
To isolate the payload trade-off, we omit the small page descriptor (8 bytes).
If each page reserves \(C\) records, then under 64-byte alignment the
approximate compressed-slot size is

\[
S_{\mathrm{slot}}(C;T,D)
 =\operatorname{align\_up}
 \left(
  12\left\lceil\frac{TD}{8}\right\rceil+4C,\ 64
 \right)\ \text{bytes}.
\]

The corresponding physical compression ratio is
\[
CR_{\mathrm{slot}}(C;T,D)
 =\frac{2TD}{S_{\mathrm{slot}}(C;T,D)}.
\]
When \(T\) and \(D\) are fixed, we write \(S(C)\) for brevity. For the
illustrative \(T=16,D=128\) configuration, \(C=16\) and \(C=32\) yield
3136-byte and 3200-byte slots, giving per-page compression ratios of
\(1.306\times\) and \(1.280\times\), respectively.

The fixed page-local capacity also constrains the choice of code width. ZipServ
uses a 3-bit code for static weights with globally indexed buffers that pool
out-of-window values across tiles~\cite{zipserv}. Our page-local format cannot
share escape capacity across pages: a 3-bit window produces a less stable
page-level escape tail, so keeping overflow risk low would require a large
record budget in every page, much of which would go unused on typical pages.
We therefore use a 4-bit Top-16 window. Its page-level escape distribution is
sufficiently stable to determine the fixed capacity from a conservative upper
bound, while avoiding excessive unused record space across pages.

\begin{figure}[tbp]
  \centering
  \IfFileExists{figures/head_page_decode_horizontal.pdf}{%
    \includegraphics[width=\columnwidth]{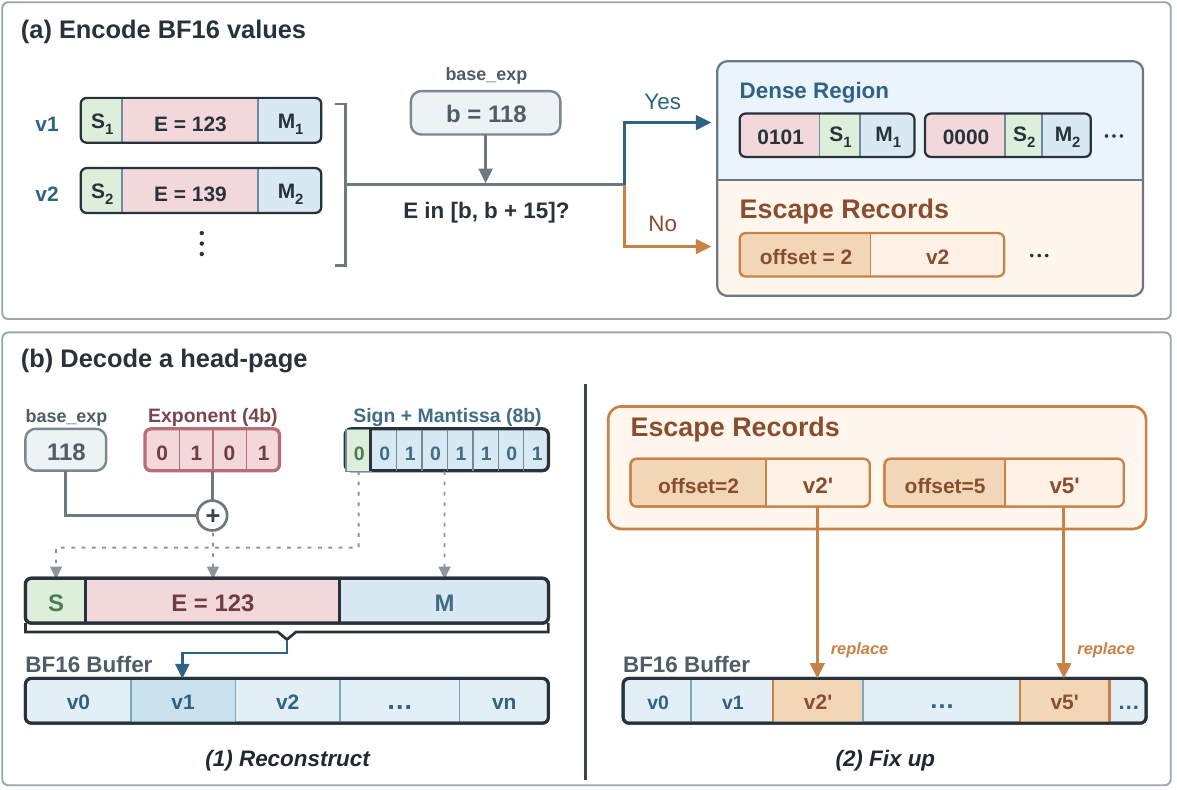}%
  }{%
    \placeholderfigure{1.8in}{Page-local encoding and decoding.}%
  }
  \caption{Page-local encoding and two-stage decoding.}
  \Description{Panel (a) shows how the exponent window determines the dense
  code and whether a BF16 value also requires an escape record. Panel (b)
  reconstructs BF16 values from the dense fields, then applies the escape
  records at their recorded positions.}
  \label{fig:head-page-decode}
\end{figure}

\runinheading{Page-Local Encoding and Decoding}
As Figure~\ref{fig:head-page-decode}(a) shows, the encoder uses the calibrated
base exponent \(b\) to check each value against the window \([b,b+15]\).
If the exponent falls within the window, the encoder stores its offset from
\(b\) as a 4-bit code in the Dense Region. Otherwise, it performs two writes.
First, it writes a placeholder code in the Dense Region. Second, it appends
the value's page-local position and original BF16 bits to the same page's
Escape Records. In both cases, the Dense Region retains the original sign
and mantissa.

Decoding reverses the encoding process in two stages, as shown in
Figure~\ref{fig:head-page-decode}(b). \textbf{First stage: Dense reconstruction.}
The decoder reads each group's 4-bit exponent codes and sign/mantissa bytes at
fixed positions. For position \(i\), it computes a temporary exponent \( \hat
e_i=b+\text{code}_i \) and concatenates \( \hat e_i \) with the preserved sign
and mantissa to form a BF16 value. For an in-window value, \( \hat e_i=e_i \),
so this stage fully reconstructs the vast majority of values; because groups
are independent, GPU threads can process the entire Dense Region in parallel
along a fixed-length path. \textbf{Second stage: Escape fix-up.}
The decoder uses each escape record's position to overwrite the corresponding
entry in the temporary buffer with the original 16-bit BF16 value. This stage
accesses only sparse exceptional positions and does not alter the uniform
Dense Region read and decode path. This organization separates the regular and
exceptional paths: the first stage applies the same fixed-length
reconstruction to every position and lets threads process groups in parallel,
while the second stage handles only sparse escape records instead of
introducing a per-value branch in the main decode loop. As a result, a small
number of out-of-window values does not change the control flow or access
pattern of dense reconstruction, which facilitates GPU decoding and avoids
warp divergence.

\subsection{KV Cache States and Storage Management}
\label{sec:design-residency}

Section~\ref{sec:design-format} gives each compressed page a fixed escape
capacity, so a small number of pages may contain more escape values than the
reserved space can hold. Although out-of-window values are rare in held-out
data, the system must still handle a page with more escape values than the
reserved capacity without truncating records. Such a page is retained in its
original BF16 representation as a \textbf{RAW} page. This requirement gives the KV Cache
two representations---raw and compressed---and requires both to be managed as
part of the page lifecycle.

\runinheading{RAW Pages and Cache Pools}
The KV Cache grows as requests produce new tokens, and completed pages may
remain available for sharing after a request finishes. At the logical level,
the system therefore exposes two cache representations. The \textbf{raw cache}
stores BF16 K/V pages for pages that are still being filled, have not yet been
submitted for compression, or cannot use the compressed format. The
\textbf{compressed cache} stores pages whose dense payload and escape records
have been generated successfully. A page is not permanently duplicated in both
representations: after a compressed page is published, its raw slot can be
reclaimed; while compression is pending or unavailable, the raw representation
remains authoritative.

At the physical level, a raw pool and a compressed pool back these two
representations. These pools are storage choices: both provide the same logical
KV pages in the same token order, while the routing layer resolves each page to
its current representation and physical slot. This indirection hides the
two-pool layout from the serving engine's upper layers, which continue to
manage logical pages through the same interface without tracking where each
page is stored. As a result, mechanisms such as scheduling and prefix sharing
need not handle compression state themselves.

\runinheading{Page States and Transitions}
To track the lifecycle of each KV page precisely, we define four page states:
\begin{description}
  \item[\textbf{OPEN}] The page is not full and can receive more tokens; its
  contents are stored as raw BF16 values.
  \item[\textbf{SEALED}] The page is full, no longer accepts tokens, and has
  not yet completed a compression attempt; it remains stored as raw BF16.
  \item[\textbf{COMPRESSED}] The compression attempt succeeds, and the page is
  read from the compressed pool.
  \item[\textbf{RAW}] The page does not use the compressed representation, for
  example because its escape records exceed the reserved capacity. It remains
  in the raw pool.
\end{description}

\begin{figure}[tbp]
  \centering
  \IfFileExists{figures/kv_lifecycle_residency_attention.pdf}{%
    \includegraphics[width=\columnwidth]{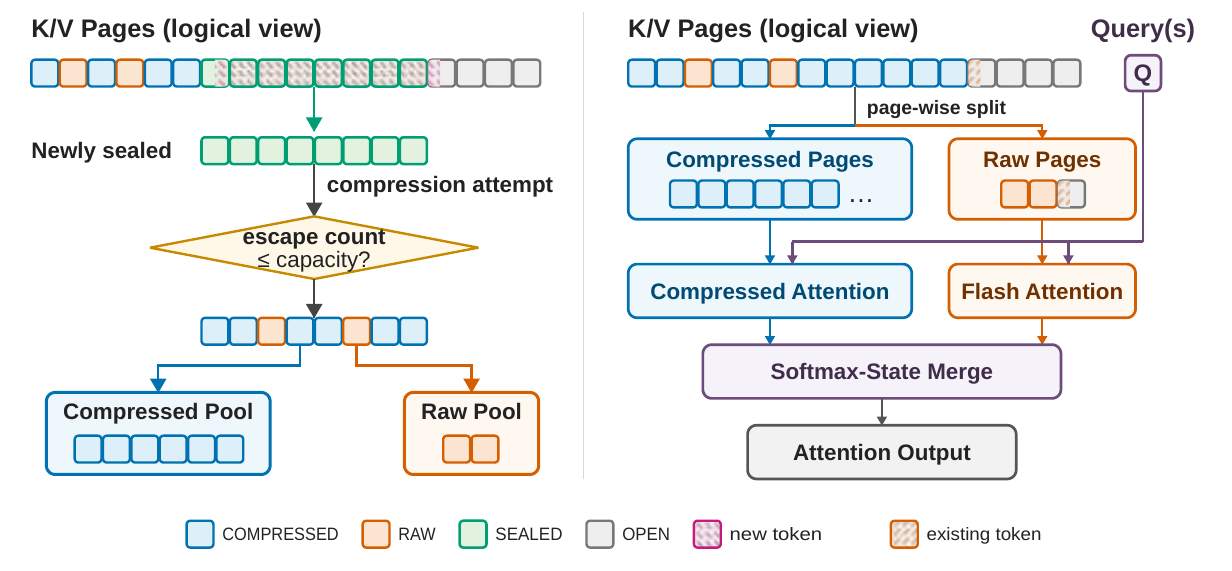}%
  }{%
    \placeholderfigure{1.8in}{RAW/COMPRESSED page production and consumption.}%
  }
  \captionsetup{justification=raggedright,singlelinecheck=false}
  \caption{RAW and COMPRESSED page production and consumption. Left (a)
  shows page-state transitions during prefill; right (b) shows
  mixed-attention data flow.}
  \Description{A two-panel figure. Panel (a) shows logical pages being sealed
  during prefill and assigned to compressed or raw storage. Panel (b) shows
  the pages routed through corresponding attention paths before their partial
  softmax states are merged.}
  \label{fig:page-centric-design}
\end{figure}

Page sealing is the transition from OPEN to SEALED. A sealed page remains
readable in the raw pool until its compressed payload is published or the
system retains the page as RAW. During decode, each request adds one token
per step and thus seals at most one growing logical page per step. Prefill can fill and
seal several pages in one pass. We use prefill as the example in
Figure~\ref{fig:page-centric-design}(a), since the single-page decode case
follows the same process. The logical sequence may already contain COMPRESSED
and RAW pages, while incoming prefill tokens fill OPEN pages and produce a set
of newly sealed pages. Each newly sealed page is tested
independently. Pages whose escape counts fit within the reserved capacity are
compressed successfully, after which their original raw slots are released for
reuse. Pages that exceed the capacity are not compressed; they remain in the
raw pool as RAW pages. Pages that are not yet full remain OPEN.

A partially filled OPEN page at the end of a request's KV sequence is called
the \textbf{open tail}. Decode appends new K/V entries to its raw BF16 slot,
and attention reads the valid entries through the raw path. The tail stays
in this representation through the step that fills it, after which the
sealed page becomes eligible for compression.

K and V may count escape values and write compressed payloads separately, but
their residency state changes as a K/V pair. If either side overflows, the
pair remains raw. Each head-page pair makes this decision independently, so a
RAW page does not prevent other layers or KV heads from using compressed
residency.

\subsection{Compressed and Mixed-Format Attention}
\label{sec:design-attention}

\runinheading{Fused Compressed Attention}
To turn the bytes saved by KV compression into an attention speedup,
decompression cannot be a separate pass that first materializes BF16 pages.
Such a pass would write the reconstructed pages to global memory and make
attention read them again, adding traffic that can offset the compression
benefit. Following ZipServ's ``load-compressed, compute-decompressed'' design~\cite{zipserv},
we therefore fuse page reconstruction with attention. This introduces
reconstruction work into the attention kernel, but Observation~2 suggests
that, in bandwidth-bound decode, much of this work can overlap with KV loading.
When the exposed overhead is small, the attention speedup can approach the
raw-to-compressed KV traffic ratio.

To realize this overlap, the kernel uses a cross-overlapped pipeline for K and
V. Loading the next K tile overlaps with decompressing the current V tile,
and loading the next V tile overlaps with decompressing the current K tile.
This organization prevents K/V loading and decompression from becoming serial
stages, allowing attention to consume each reconstructed tile as soon as it is
ready.

\runinheading{Mixed-Format Attention}
During decode, a query must attend to every valid KV entry in its visible
context. These entries can span compressed pages, an open raw tail, and
sealed pages that remained raw after overflow. We route each representation
through its own attention path to keep the inner loops specialized for
their respective data layouts, as shown in
Figure~\ref{fig:page-centric-design}(b). Compressed pages use the kernel
described above, while raw pages follow FlashAttention. Both paths receive
the same query.
The split is logical rather than physical: in a paged attention
system~\cite{pagedattention}, the paging mechanism resolves each logical page
to its physical slot, so the two sets need not be contiguous, moved, or laid
out together. Their logical
positions remain unchanged.

Each path normalizes its partial output over its own pages and retains the
log-sum-exp (LSE) of their attention scores. The LSE determines each path's
weight in the global softmax. Even a short raw tail can carry substantial
weight when its tokens have high scores.
FA2 combines split-KV partials using LSE rescaling~\cite{flashattention2}.
We apply the same reduction across the raw and compressed paths. Let
\((O_C,L_C)\) and \((O_R,L_R)\) denote their aggregate outputs and LSE states.
With \(m=\max(L_C,L_R)\), the merged output is

\[
O=\frac{e^{L_C-m}O_C+e^{L_R-m}O_R}
        {e^{L_C-m}+e^{L_R-m}}.
\]

Subtracting \(m\) keeps both exponentials at most one while preserving their
ratio. The denominator normalizes over the full context because the two paths
cover disjoint pages. The merge uses only the partial outputs and LSEs, so the
pages can stay in separate pools.

\subsection{Offline Calibration and Deployment Configuration}
\label{sec:design-calibration}

Offline calibration produces the deployment artifact used by the online
runtime. It specifies the exponent windows and the fixed escape-record
capacities \(C_K\) and \(C_V\) for compressed K and V slots. It also records
the overflow estimates used to provision the RAW pool.

We calibrate exponent windows using BF16 KV cache samples, keeping calibration
and validation traces separate. For each layer, KV head, and K/V type,
calibration observations determine the contiguous 16-exponent
interval with the highest coverage. We then count escape records per head-page
on held-out validation traces to guide capacity selection.

\begin{figure}[tbp]
  \centering
  \IfFileExists{figures/head_page_fallback_quantiles.pdf}{%
    \includegraphics[width=\columnwidth]{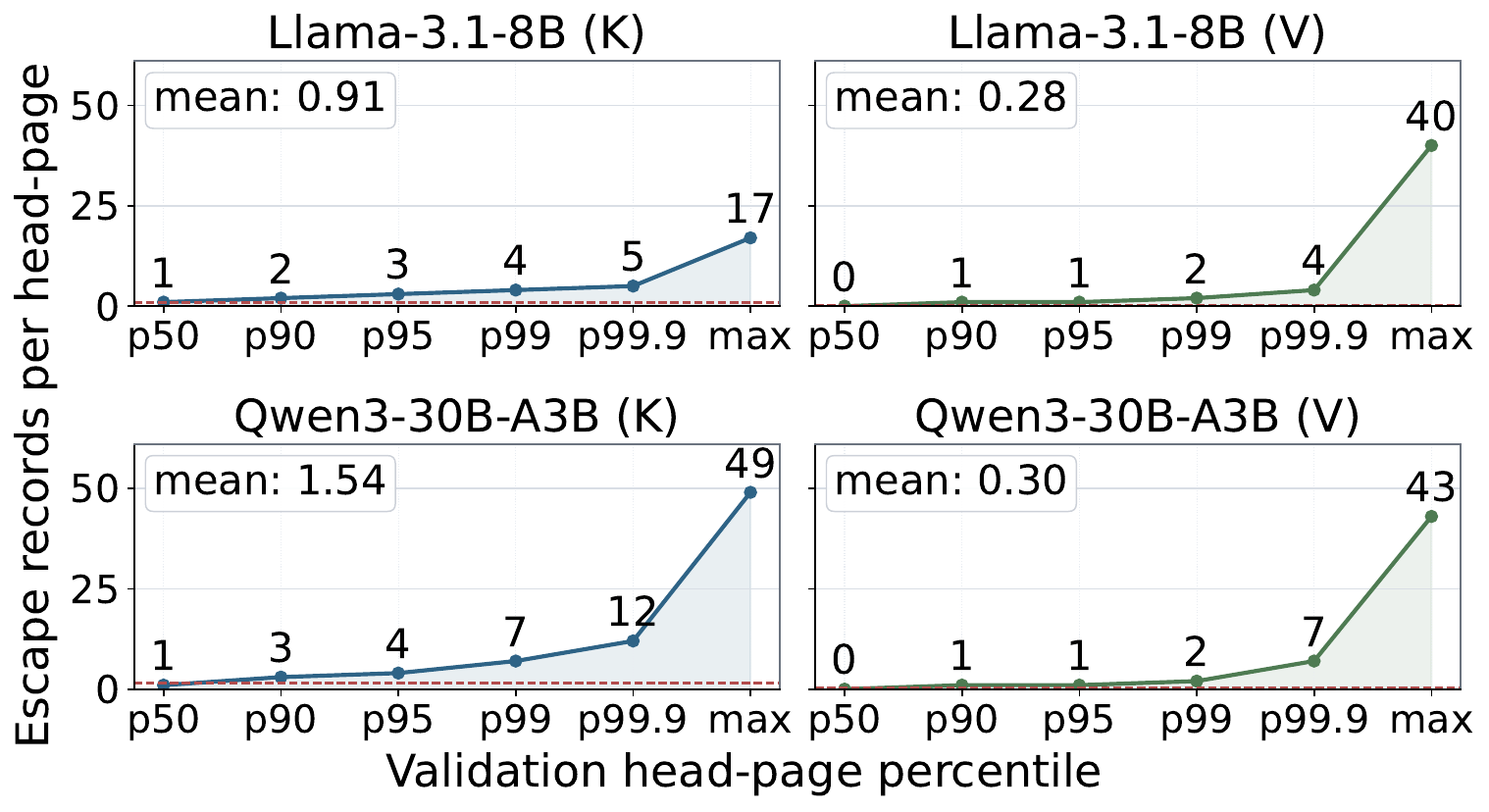}%
  }{%
    \placeholderfigure{1.45in}{Head-page escape-count tail (figure pending).}%
  }
  \caption{Head-page escape-record quantiles for capacity selection.}
  \Description{Four panels show escape-record quantiles for Keys and Values
  in Llama-3.1-8B-Instruct and Qwen3-30B-A3B-Instruct-2507, highlighting the
  difference between the common case and the tail.}
  \label{fig:escape-tail}
\end{figure}

Choosing the fixed escape capacities \(C_K\) and \(C_V\) is not trivial.
Figure~\ref{fig:escape-tail} shows a pronounced gap between typical and
worst-case pages: V pages contain only 0.28--0.30 escape records on average,
yet their observed maxima reach 40--43; the K-side maximum reaches 49.
Sizing every compressed slot for such rare extremes would waste space,
whereas capacities that are too small would send more pages to the RAW pool.
Byte alignment can make several record capacities occupy the same slot size.
Within that size, the alignment padding can hold additional escape records
without increasing the slot footprint.

We use a heuristic to choose the escape capacities and size the RAW pool.
We select the smallest aligned capacities for K and V that keep their
respective overflow rates on held-out pages at or below a preset threshold,
then use the estimated page-pair overflow rate at those capacities to reserve
RAW slots for the expected overflow pages. The pool also reserves slots for
the open tails of concurrent requests.

The size of the RAW pool affects serving stability. When the compressed pool
fills, \sys keeps newly sealed pages uncompressed in their RAW slots so
decoding can continue. We call this transition \textbf{RAW fallback}.
If the RAW pool fills, decoding can stop even while GPU memory is still
available. To reduce the risk of such an early stop, we add a protected RAW
reserve to the allocation above, with a default floor of 1\% of the logical
head-pages. This reserve is small enough to have little effect on the overall
compression ratio, yet substantially exceeds the expected overflow rate and
gives the system room to absorb bursts.

\section{Implementation}
\label{sec:implementation}

We extend FlashAttention-2 with
page-local reconstruction and mixed raw/compressed execution. The serving
system is built on vLLM: we change how KV pages are stored physically while
preserving the logical page interface exposed to the rest of the serving
stack. Upper layers therefore need not track whether a page is raw or
compressed. The GPU worker maintains a per-request index of raw and
compressed slots and updates it as pages change state. Before attention,
the worker uses this index to build the page tables for the two paths.

The implementation comprises approximately 15K lines of code: about 8K lines
of Python, 5K lines of CUDA, and 2K lines of C++. The code covers the online
KV-page lifecycle, lossless codec, residency management, and fused attention
path evaluated in this paper.

\section{Evaluation}
\label{sec:evaluation}

We evaluate \sys in four aspects: kernel performance, end-to-end serving
performance, numerical correctness and task fidelity, and calibration
robustness.

\runinheading{Testbed}
All experiments are conducted on two GPU platforms. The consumer-grade
platform contains 4x 24-GB NVIDIA RTX 4090 GPUs and a 24-core Intel
Xeon Sapphire Rapids CPU, and uses CUDA 12.9. The datacenter platform contains
4x 48-GB NVIDIA L40S GPUs and an AMD EPYC 7543 host exposing 46 CPUs,
and uses CUDA 13.0.

\runinheading{Baselines}
At the kernel level, we assess compression and attention against six lossless
methods. These include nvCOMP Bitcomp, DietGPU, and
DFloat11~\cite{nvcomp,dietgpu,dlf}, as well as SplitZip, IBP-GPU, and
ZipServ~\cite{splitzip,kamath2026lossless,zipserv}. Raw BF16 copying and
FlashAttention-2~\cite{flashattention2} serve as uncompressed references. For
end-to-end serving, vLLM~\cite{pagedattention} is the unmodified baseline with
its original FlashAttention-2 path. This comparison tests
whether kernel-level gains carry over within the same serving framework.

\runinheading{Models and datasets}
For end-to-end serving, we run Qwen3-4B-Instruct-2507 and
Llama-3.1-8B-Instruct on four RTX 4090 GPUs, and
Mistral-Small-3.2-24B-Instruct-2506 and Qwen3-30B-A3B-Instruct-2507 on four L40S GPUs.
We use PG-19~\cite{rae2020compressive} for offline calibration and
LongBench-v1~\cite{bai2024longbench} for task-level evaluation. To test
calibration across domains, we also use MATH~\cite{hendrycks2021math} and
CodeSearchNet~\cite{husain2019codesearchnet} alongside PG-19 and LongBench-v1.

We time kernels for 1,000 iterations after 100 warm-up iterations.
Serving uses \(\mathrm{PP}=4\). Each configuration has one warm-up run
and three measured runs. Both serving paths use the same inputs, per-GPU
budget, and context settings. Task fidelity uses the models' native positional
encoding.

\subsection{Kernel-Level Performance}
\label{sec:evaluation-kernel}

\begin{table}[t]
  \centering
  \caption{BF16 KV encode/decode performance on RTX 4090.}
  \label{tab:eval-codec}
  \scriptsize
  \setlength{\tabcolsep}{2.4pt}
  \resizebox{0.9\columnwidth}{!}{%
  \begin{tabular}{@{}lrrr@{}}
    \toprule
    Method & Ratio & Encode (GB/s) & Decode (GB/s) \\
    \midrule
    Raw BF16 copy & 1.000 & 502.8 $\pm$ 1.1 & 502.8 $\pm$ 1.1 \\
    nvCOMP Bitcomp & 1.092 & 206.1 $\pm$ 3.0 & 83.0 $\pm$ 0.8 \\
    DietGPU & \textbf{1.409} & 203.9 $\pm$ 1.3 & 224.6 $\pm$ 1.1 \\
    DFloat11 & 1.402 & 0.0107~(CPU) & 450.7 $\pm$ 3.6 \\
    SplitZip & 1.313 & 289.8 $\pm$ 7.6 & 471.1 $\pm$ 1.6 \\
    IBP-GPU & 1.127 & 45.6 $\pm$ 0.1 & 184.8 $\pm$ 0.3 \\
    ZipServ & 1.312 & 0.1106~(CPU) & 511.6 $\pm$ 2.1 \\
    \sys & 1.306 & \textbf{542.4 $\pm$ 1.3} & \textbf{532.0 $\pm$ 1.4} \\
    \bottomrule
  \end{tabular}%
  }
\end{table}

\begin{figure}[tbp]
  \centering
  \includegraphics[width=\columnwidth]{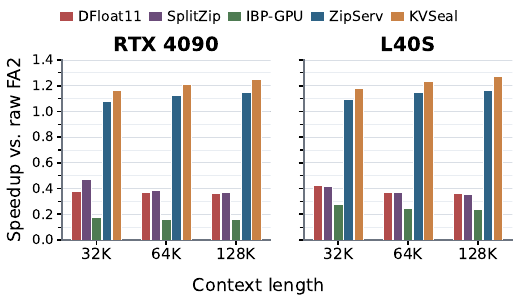}
  \caption{Relative attention speedup at 32K--128K contexts.}
  \Description{Two panels show relative attention speedup for five methods at
  32K, 64K, and 128K contexts on an RTX 4090 and an L40S.}
  \label{fig:eval-attention}
\end{figure}

\begin{figure}[!b]
  \centering
  \includegraphics[width=\columnwidth]{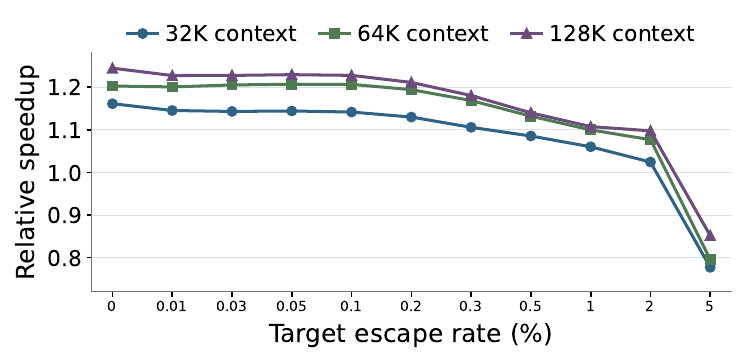}
  \caption{Attention speedup under controlled escape rates.}
  \Description{Relative attention speedup versus target escape rate for 32K,
  64K, and 128K contexts. Circles, squares, and triangles distinguish the
  contexts. The tested escape rates are shown at equal horizontal spacing, and
  the dashed line marks raw FA2.}
  \label{fig:eval-escape-sweep}
\end{figure}

\begin{figure*}[!t]
  \centering
  \includegraphics[width=\textwidth]{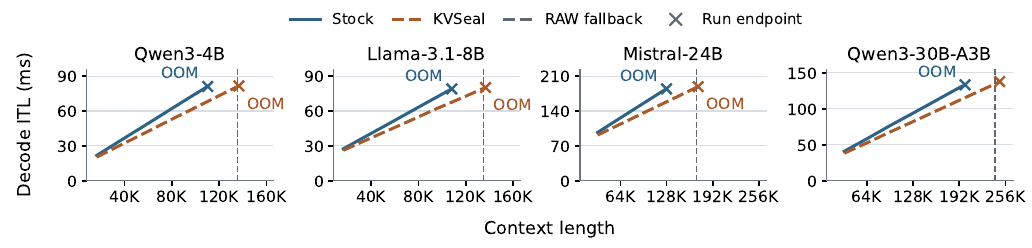}
  \caption{Decode latency and capacity at \(\mathrm{batch}=4\). RAW fallback indicates compressed pool exhaustion.}
  \Description{Four panels compare Stock and KVSeal decode latency versus
  per-request KV length. Dashed vertical lines mark RAW fallback, offset to
  the left of their measured positions. Crosses mark run endpoints, with OOM
  labels for KV capacity failures. The later Mistral KVSeal windows use an
  extended position limit.}
  \label{fig:eval-context}
\end{figure*}

\runinheading{Encode/decode performance}
Using BF16 KV from a 16K-token context, we benchmark encoding and decoding
on an RTX 4090 and compare the storage required by each method. To express
the storage savings as a compression ratio, we divide the original BF16
size by the size of the encoded representation. Table~\ref{tab:eval-codec}
reports these ratios alongside throughput, computed as the original BF16
size divided by the elapsed time for each operation.
DietGPU and DFloat11 achieve the highest ratios,
at \(1.409\times\) and \(1.402\times\), respectively. \sys reaches
\(1.306\times\) with \(C_K=C_V=16\), including the fixed escape reserve.
ZipServ's 3-bit exponent code gives a theoretical compression ratio of
\(16/11 \approx 1.455\times\), which drops to \(1.312\times\) on this input
once full BF16 escape values and tile offsets are included~\cite{zipserv}.
SplitZip uses a similar 4-bit exponent format to \sys and achieves a
slightly higher ratio of \(1.313\times\)~\cite{splitzip}. Compared with these
two codecs, \sys trades a small amount of compression for predictable
storage and regular decoding.

\sys leads the tested compressors in both encode and decode throughput,
at 542.4 and 532.0~GB/s. Encoding is \(2.7\times\) faster
than DietGPU and \(1.9\times\) faster than SplitZip; decoding is 4\% faster
than ZipServ and 18\% faster than DFloat11. Both throughputs also exceed the
502.8~GB/s raw BF16 copy reference in this microbenchmark. This is consistent
with lightweight bit reconstruction whose performance is dominated by memory
traffic rather than arithmetic in this setting.

\runinheading{Compressed attention}
We compare two ways of consuming compressed KV in single-query decode
attention on the RTX 4090 and L40S, with 32K, 64K, and 128K KV histories.
KVSeal and the ZipServ-style baseline use a fused path, reconstructing KV as
part of the attention operator, following the fused load--decompress--compute
organization described by ZipServ~\cite{zipserv}. DFloat11, SplitZip, and
IBP-GPU first decode the compressed KV into a BF16 workspace and then call
regular FA2. Their reported times include both operations. For SplitZip,
this follows its original receiver-side
reconstruction path~\cite{splitzip}. A fused serving path using its shared
escape stream across tensor chunks would require additional indexing to
support independent allocation and reclamation of resident KV pages.

Figure~\ref{fig:eval-attention} shows that KVSeal's fused path improves as the
context grows, from \(1.16\times\) to \(1.25\times\) on the RTX 4090 and from
\(1.18\times\) to \(1.27\times\) on the L40S between 32K and 128K. The peak
speedup is close to the roughly \(1.3\times\) raw-to-resident byte ratio
reported in Table~\ref{tab:eval-codec}, indicating that the saved KV traffic is
translated into attention speed when reconstruction remains lightweight.

\begin{samepage}
The ZipServ-style path reaches
only \(1.15\times\) and \(1.17\times\) at 128K, despite having a similar nominal
compression ratio. This gap is consistent with a more involved decode and
metadata path: extra instructions, control flow, and register use can compete
with the attention computation inside the fused operator. The three
materialization paths run at \(0.16\text{--}0.37\times\) the speed of regular
FA2 at 128K across the two GPUs. Writing the recovered BF16 KV to memory
and reading it again for attention adds traffic on top of the decoder's own
work. Together, these results show that translating KV byte
savings into faster decode attention depends on avoiding the BF16 round trip
and keeping reconstruction lightweight inside the attention loop.
\end{samepage}

\runinheading{Escape-rate sensitivity}
To probe the kernel's sensitivity to escape values, we construct synthetic KV
inputs with controlled escape rates and measure attention speed relative to raw
FA2 at fixed context lengths. At 128K on the RTX 4090, speedup decreases from
\(1.23\times\) at a target escape rate of 0.1\% to \(1.11\times\) at 1\%,
then drops below parity to \(0.85\times\) at 5\%
(Figure~\ref{fig:eval-escape-sweep}). In the profiled workloads, however, the
cumulative prompt-KV escape rate stays below 0.08\% across all four models
(Figure~\ref{fig:eval-escape-context}). This puts the observed rates in the
range where the kernel maintains a stable speedup in the sweep.

\subsection{End-to-End Serving Performance}
\label{sec:evaluation-serving}

\runinheading{Decode latency and KV capacity}
We evaluate decode latency and maximum context length at \(\mathrm{PP}=4\)
and \(\mathrm{batch}=4\). The setup pairs Qwen3-4B and Llama-3.1-8B
with four RTX 4090 GPUs, and Mistral-24B and Qwen3-30B-A3B with four L40S GPUs.
Offline calibration selects the K and V escape capacities for each model.
Stock vLLM and \sys use the same per-GPU decode KV budget. We measure decode
latency as the context grows from 16K tokens per request toward the KV
capacity limit. Figure~\ref{fig:eval-context} reports median inter-token
latency (ITL) at the measured context lengths.

As the context grows, attention reads more KV data and accounts for a larger
share of decode latency, giving \sys more opportunity to speed up decoding
by reducing KV traffic. At the shared measurement point of 96K tokens,
this yields a \(1.13\times\) to \(1.19\times\) speedup across the four models.
Moreover, \sys reaches longer contexts than stock vLLM in our runs for all
four models. For example, under the same KV budget, the capacity limit for
Llama-3.1-8B rises from about 108K to 137K tokens per request.
Across four models, \sys increases capacity by
24.0--26.4\%, close to the roughly 30\% theoretical capacity gain implied by
the compression format, showing that most of the space saved by compression
translates into room for longer contexts.

Figure~\ref{fig:eval-context} also shows the two stages of KV capacity exhaustion
described in \S~\ref{sec:design-calibration}. The dashed lines indicate the
onset of RAW fallback when compressed slots run out. Across the four models,
the observed fallback phase spans roughly 160--500 tokens per request,
under 1\% of the final context length. This fallback occurs at the edge of
usable KV capacity, leaving ordinary decoding unaffected while compressed
slots remain available.

\begin{figure}[tbp]
  \centering
  \includegraphics[width=\columnwidth]{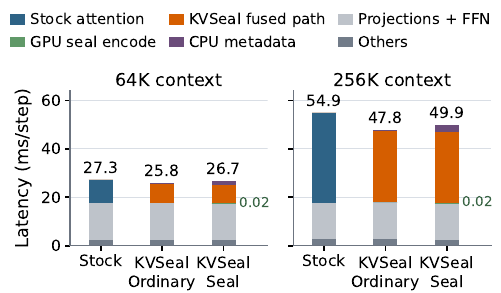}
  \caption{Llama-3.1-8B decode breakdown on 4090.}
  \Description{Two side-by-side panels share a latency scale. Each shows
  three stacked bars for stock vLLM, an ordinary KVSeal step, and a KVSeal
  sealing step. The bottom segments show other runtime work and projections
  plus FFN. Above them, blue marks stock attention and orange marks the
  KVSeal fused path. A green line below the orange segment in each sealing
  bar marks GPU encoding, labeled 0.02 ms. Purple shows exposed CPU
  compression metadata time in both KVSeal bars.}
  \label{fig:eval-decode-breakdown}
\end{figure}

\runinheading{Decode breakdown}
To understand where \sys saves time and where it adds overhead, we profile
Llama-3.1-8B on four RTX 4090 GPUs at \(\mathrm{batch}=1\) with 64K and 256K
contexts. As Figure~\ref{fig:eval-decode-breakdown} shows, projection and FFN
times are nearly unchanged, while \sys reduces attention time from 9.65 to
7.84~ms at 64K and from 37.13 to 29.44~ms at 256K, reductions of about 19\%
and 21\%, respectively. As attention's share of stock decode rises from
35\% to 68\% with context length, the speedup for ordinary decode steps grows
from \(1.06\times\) at 64K to \(1.15\times\) at 256K.

Sealing adds two costs to a decode step. First, a lightweight GPU kernel
compresses newly filled pages after the model forward pass. Threads pack
fixed-width exponent codes in parallel and write the few out-of-window
values to the page-local escape region. This keeps encoding to about
0.02~ms per sealing step at both context lengths. Second, sealing a full page
requires CPU metadata updates to record its new state and slot ownership.
Because the next decode step needs only some of these updates, \sys can defer
the rest until after submitting that step's CUDA Graph, allowing the deferred
updates to overlap GPU execution. In total,
sealing adds about 1--2~ms relative to an ordinary step. With 16-token
pages, a single request seals one page every 16 decode steps, amortizing
the overhead across those steps.

\begin{figure}[!b]
  \centering
  \includegraphics[width=\columnwidth]{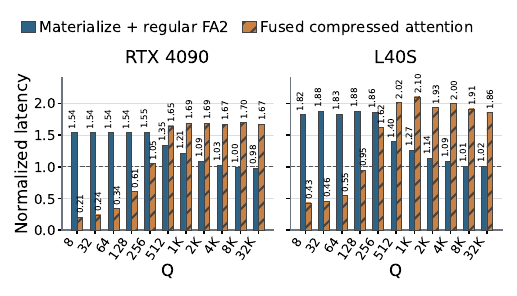}
  \caption{Prefill overhead with 32K cached KV (1.0 = FA2).}
  \Description{Two panels compare normalized execution time for fused
  compressed attention and materialization followed by regular FA2 on an RTX
  4090 and an L40S; paged FA2 is the 1.0 reference.}
  \label{fig:eval-prefill}
\end{figure}

\runinheading{Prefill overhead analysis}
To assess how compressed KV affects prefill attention, we vary the number of
new query tokens while holding the cached KV history at 32K. As
Figure~\ref{fig:eval-prefill} shows, materializing the KV once before calling
FA2 takes \(1.54\text{--}1.82\times\) the stock FA2 time for an 8-token query
across the two GPUs. With 32K query tokens, this ratio falls to
\(0.98\text{--}1.02\times\) as the one-time reconstruction cost becomes
small relative to attention computation. Fused compressed attention takes
only \(0.21\text{--}0.43\times\) the stock time at 8 tokens by reducing KV
traffic, but it reconstructs the same KV for more query tiles as the query
grows, reaching \(1.67\text{--}1.86\times\) the stock time at 32K tokens.
The two paths cross between 256 and 512 query tokens on both GPUs in this
sweep. A query-length threshold can therefore use fusion for short queries
and switch to materialization for longer ones, retaining the short-query
speedup and keeping long-query latency close to stock FA2.

\subsection{Correctness and Task Fidelity}
\label{sec:evaluation-fidelity}

Our losslessness guarantee is defined at the KV representation level. The
decoder reconstructs every compressed page to exactly the original BF16 bit
pattern, including values stored in escape records. In this sense, \sys is
bit-exact: compression does not alter the KV values consumed by attention.

Bitwise equivalence of the final attention output is a separate property.
FlashAttention partitions the KV sequence into multiple splits, computes
partial online-softmax statistics, and merges them using a floating-point
reduction. The resulting reduction order depends on both the split schedule
and the partition boundaries. Since floating-point addition is not
associative, two attention implementations that consume identical K/V bits
can still produce small differences in intermediate statistics and output
values. Our compressed path uses a different split configuration from the
Raw path, so these numerical differences may occasionally propagate to
different generated tokens at long context lengths. This behavior is a
consequence of the reduction schedule, rather than an error in lossless KV
reconstruction.

We evaluate task fidelity for Llama-3.1-8B and Mistral-24B on six
LongBench-v1 tasks with 200 held-out examples per task. The results show that the first generated
token matches between stock vLLM and \sys for over 98\% of requests for both
models, and the full output token sequences match for 70.1\% of Llama requests
and 81.1\% of Mistral requests. As Table~\ref{tab:eval-fidelity} shows,
the two paths score within one point of each other on every task for both
models, with \sys scoring higher on some tasks and lower on others. The
six-task mean drops by 0.21 points for Llama and 0.09 points for Mistral.
These small score differences suggest that changes in reduction order have
little effect on task accuracy in this evaluation.

\begin{table}[t]
  \centering
  \caption{LongBench-v1 scores for Stock and \sys.}
  \label{tab:eval-fidelity}
  \scriptsize
  \setlength{\tabcolsep}{2pt}
  \renewcommand{\arraystretch}{0.95}
  \resizebox{\columnwidth}{!}{%
  \begin{tabular}{@{}lrrrrrrr@{}}
    \toprule
    Path & LCC & SAM & TQA & QMS & NQA & QAS & Avg. \\
    \midrule
    \multicolumn{8}{@{}l@{}}{\textbf{Llama-3.1-8B}} \\
    Stock & 63.51 & 43.44 & 91.65 & 25.26 & 29.57 & 44.88 & 49.72 \\
    \sys  & 63.28 & 43.64 & 90.98 & 24.88 & 29.52 & 44.74 & 49.51 \\
    \midrule
    \multicolumn{8}{@{}l@{}}{\textbf{Mistral-Small-3.2-24B}} \\
    Stock & 70.16 & 34.28 & 84.17 & 25.16 & 35.84 & 50.70 & 50.05 \\
    \sys  & 70.30 & 33.69 & 83.96 & 25.11 & 35.95 & 50.73 & 49.96 \\
    \bottomrule
  \end{tabular}%
  }
\end{table}

\begin{figure}[!b]
  \centering
  \includegraphics[width=\columnwidth]{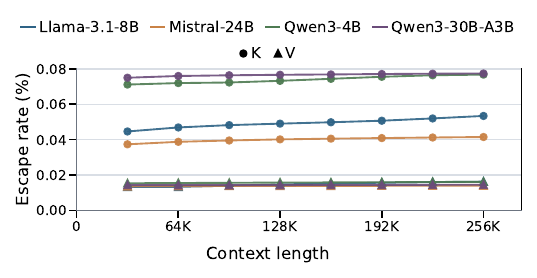}
  \caption{Cumulative prompt-KV escape rates versus context length.}
  \Description{A single plot shows cumulative K and V escape rates for four
  models. Color identifies the model, and circle and triangle markers identify
  K and V, respectively. The y-axis starts at zero, and the x-axis is marked
  at 64K intervals. Values are aggregated across layers and KV heads from one
  longest-context prompt per model at approximately 32K-token increments;
  generated-token KV is excluded.}
  \label{fig:eval-escape-context}
\end{figure}

\subsection{Calibration Robustness}
\label{sec:evaluation-calibration}

The effectiveness of our fixed exponent window rests on two assumptions:
(1) escape rates remain stable as context length grows, and (2) calibration
data remain representative of the inputs encountered during serving. If
either assumption fails, more head-pages may exceed their escape capacity
and remain RAW.

To test assumption (1), we track the cumulative value-level escape rate for
four models as the prompt grows to approximately 256K tokens.
Figure~\ref{fig:eval-escape-context} shows that the cumulative K and V escape
rates stay below 0.08\% throughout the measured range. At 256K, the four
models span approximately 0.041--0.077\% for K and 0.014--0.016\% for V.
The largest absolute increase from 32K to 256K is only 0.009\%, for K in
Llama-3.1-8B. This stability suggests that growing the context is unlikely
to trigger a surge in overflowing pages that would rapidly consume the
reserved RAW slots.

To test assumption (2), we calibrate separate windows for each of four domains
on Llama-3.1-8B and apply them to KV from held-out documents in each domain.
We use the K/V head-page pair fallback rate because escape values can cluster
within a page, making the value-level escape rate a poor predictor of how
many pages stay RAW. As Table~\ref{tab:eval-calibration} shows, even the highest
cross-domain fallback rate is only about 0.05\%, and nine of the twelve cross-domain pairs stay
below 0.01\%. These results provide strong evidence that calibrated windows
transfer across the tested domains while keeping page fallback rare.

\begin{table}[t]
  \centering
  \caption{Cross-domain page-pair fallback rates for Llama-3.1-8B.}
  \label{tab:eval-calibration}
  \small
  \setlength{\tabcolsep}{1.5pt}
  \begin{tabular*}{\columnwidth}{@{}l@{\extracolsep{\fill}}rrrr@{}}
    \toprule
    & \multicolumn{4}{c}{Calibration domain} \\
    \cmidrule(lr){2-5}
    Evaluation domain & LB & PG-19 & MATH & CSN \\
    \midrule
    LongBench-v1  & \(<\)0.001\% & 0.008\%      & 0.048\%      & 0.007\% \\
    PG-19         & \(<\)0.001\% & \(<\)0.001\% & 0.029\%      & \(<\)0.001\% \\
    MATH          & \(<\)0.001\% & \(<\)0.001\% & \(<\)0.001\% & \(<\)0.001\% \\
    CodeSearchNet & 0.001\%      & 0.004\%      & 0.012\%      & \(<\)0.001\% \\
    \bottomrule
  \end{tabular*}
  \Description{A four-by-four matrix of measured page-pair fallback rates.
  Rows are held-out evaluation domains and columns are calibration domains.
  LB is LongBench-v1 and CSN is CodeSearchNet.}
\end{table}

\section{Discussion and Limitations}
\label{sec:discussion}

\runinheading{Compatibility with serving mechanisms}
\sys changes the physical representation and allocation pools of KV pages but
preserves their logical page interface. Higher-level serving mechanisms need
not track whether a page is raw or compressed. Prefix sharing can reuse
compressed pages, while offload and prefill--decode disaggregation can move
them with their metadata. Such integration must preserve page identity and
reference lifetime. Our prototype uses vLLM's paged allocation and partial
prefix-management support; interactions with eviction, preemption, and
asynchronous scheduling remain to be evaluated.

\runinheading{Relationship to other KV-cache optimizations}
Sparsity, retrieval, and eviction determine which KV entries attention
accesses; \sys instead changes how retained pages are represented. These
approaches are therefore complementary. Quantization could also be combined
with \sys, but it changes the numerical representation before lossless coding.
Such a combination would need to find redundancy in the new format and state
whether losslessness refers to the original BF16 or quantized values.

Our prototype evaluates MHA and GQA. Other attention states, such as MLA
latents and recurrent states in linear attention, may also contain compressible
BF16 patterns. Extending page-local compression to them would require adapting
to their different layouts and update rules.

\runinheading{Hardware and prefill limitations}
\sys benefits most when saved KV traffic outweighs reconstruction work. This
balance may be less favorable on accelerators with more memory bandwidth
relative to compute: KV loads become cheaper, while recovering compressed
values still incurs instructions. On such hardware, compression may save
capacity without accelerating attention. A lighter reconstruction path could
broaden its performance benefit.

Prefill poses a related limitation. With many new queries, attention can be
compute-bound, so KV loading may not hide reconstruction.
Our current policy chooses between compressed attention and one-time
materialization followed by FA2 according to query and context lengths, to
avoid a substantial prefill regression. The crossover in
Figure~\ref{fig:eval-prefill} was measured with a fixed KV history. For other
history lengths or GPUs, the switch point may shift as the cost of reading
and reconstructing KV changes. Choosing the materialization path also
requires budgeting for the temporary BF16 workspace. A more efficient
multi-token reconstruction path remains future work for reducing time to
first token.

\section{Related Work}
\label{sec:related-work}

\runinheading{Approximate KV Reduction}
KV-cache traces reveal reuse patterns that inform eviction~\cite{wang2025kvcachewild}. MInference, Quest, Squeezed Attention,
RetrievalAttention, and SolidAttention use sparse or retrieval-based attention
to select fewer tokens
~\cite{minference,quest,squeezedattention,retrievalattention,solidattention}.
SmallThinker reduces KV storage for local inference by alternating global
NoPE layers with sliding-window RoPE layers~\cite{smallthinker}.
Quantization methods such as KIVI, KVQuant, and GEAR reduce the precision of
KV values, while CacheGen, TurboQuant, VecInfer, and KV Cache Transform Coding
explore compressed representations for serving or storage
~\cite{kivi,kvquant,gear,liu2024cachegen,turboquant,vecinfer,
kv-transform-coding}. These methods can provide substantial memory or compute
savings, but may introduce approximation error or change the set of historical
information available to attention~\cite{liu2025fundamental}. KIVI also
shows how attention can consume quantized groups alongside a full-precision
residual cache~\cite{kivi}. In \sys, open pages and pages that exceed their
escape capacity remain raw, while compressed pages preserve the original
BF16 bits.

Other systems combine approximate representations with mechanisms for
recovering or reusing information. KeepKV retains a complete external cache
to provide periodic losslessness, VeriCache verifies compressed generation
against a full KV copy, and CacheBlend reuses KV states from retrieved
contexts~\cite{tian2026keepkv,yao2026vericache,yao2025cacheblend}. Their
objectives differ from ours: \sys preserves every BF16 value in the resident
cache and reduces the cost of storing and consuming pages that attention still
needs to access.
\runinheading{Lossless Compression}
ZipNN and related GPU codecs demonstrate that neural-network tensors can be
compressed and decoded in parallel~\cite{zipnn,dietgpu,recoil}. Dynamic-length
float coding and transform coding extend this idea to model components and KV
transfers~\cite{dlf,kv-transform-coding}. SplitZip uses 4-bit exponent codes
and sparse corrections for KV transfers between prefill and decode
workers~\cite{splitzip}. ZipServ fuses decompression of static model weights
with GEMM~\cite{zipserv}. Kamath et al.~\cite{kamath2026lossless}
exploit bit patterns shared across tensors in CPU--GPU offload.
\sys applies exponent coding and fused reconstruction to resident KV
pages. Its bounded escape region allows independent page recovery and
reclamation, and its K/V layouts let decode attention read the pages in
their stored formats.
\runinheading{Inference Systems}
PowerInfer and NeuronInfer distribute LLM computation across CPU and GPU on
consumer PCs and across CPU and NPU on smartphones,
respectively~\cite{powerinfer,neuroninfer}.
FlashAttention and FlashAttention-2 reduce attention I/O through tiling and
online softmax~\cite{flashattention,flashattention2}. PagedAttention
introduces block-based KV-cache management for serving
~\cite{pagedattention}, while FlashInfer provides an extensible attention
engine for heterogeneous serving workloads~\cite{flashinfer}. \sys builds on
tiled attention and paging, adding page-local compression to change how KV
pages are stored, paired K/V publication to manage their residency, and
mixed-format dispatch to consume them.

\section{Conclusion}
\label{sec:conclusion}

Long-context inference makes the KV Cache a capacity and decode-bandwidth
bottleneck. Approximate reduction can impair difficult tasks; \sys instead
preserves every BF16 K/V value through three coordinated designs. Page-local
Top-16 coding uses fixed slots and sparse escape records for exact recovery.
Lifecycle management seals and classifies pages, retaining raw and compressed
representations in separate pools behind one logical interface. Mixed
attention reconstructs compressed values on chip, processes both page types,
and merges their results without a full BF16 cache. Together, these designs
make lossless compression a resident KV-cache representation for online serving.

The representation delivers up to \(1.3\times\) physical KV compression
with bitwise round-trip correctness. Compressed attention reaches up to
\(1.27\times\) the speed of FA2, and \sys achieves up to a \(1.19\times\)
end-to-end decode speedup over stock vLLM. The codec also leads the
compared baselines in encode and decode throughput. Its lossless
representation can complement sparsity, retrieval, and eviction in
long-context serving.

\bibliographystyle{ACM-Reference-Format}
\bibliography{references}

\end{document}